# High-throughput high-resolution multi-scale three dimensional laser printing

Yuanxin Tan,[1,2] Wei Chu,[1,*] Peng Wang,[1,3] Wenbo Li,[1,2,4] Jia Qi,[1,2] Jian Xu,[5,6] Ya Cheng[1,5,6,7,*]

*Corresponding Author: E-mail: chuwei0818@qq.com   ya.cheng@siom.ac.cn

[1]State Key Laboratory of High Field Laser Physics, Shanghai Institute of Optics and Fine Mechanics, Chinese Academy of Sciences, Shanghai 201800, China
[2]University of Chinese Academy of Sciences, Beijing 100049, China
[3]School of Physics Science and Engineering, Tongji University, Shanghai 200092, China
[4]School of Physical Science and Technology, Shanghai Tech University, Shanghai 200031, China.
[5]State Key Laboratory of Precision Spectroscopy, School of Physics and Materials Science, East China Normal University, Shanghai 200062, China
[6]XXL - The Extreme Optoelectromechanics Laboratory, School of Physics and Materials Science, East China Normal University, Shanghai 200241, China
[7]Collaborative Innovation Center of Extreme Optics, Shanxi University, Taiyuan, Shanxi 030006, China

**Abstract**: Throughput and resolution are both of critical importance for modern manufacturing, however, they are usually contradictory with each other. Here, a high-throughput high-resolution three dimensional (3D) printing is demonstrated by incorporating a simultaneous spatiotemporal focusing (SSTF) scheme in two-photon polymerization (TPP). Remarkably, the SSTF can ensure generation of a spherical focal spot of which the size depends linearly on the laser power. Thus, the resolution can be continuously adjusted from sub-10 μm to ~40 μm by only increasing the laser power. A multi-scale 3D structure is fabricated using this technique, for which the regions of coarse and fine feature sizes are produced at high and low resolutions, respectively.

# 1. Introduction

Femtosecond laser two-photon polymerization has enabled 3D laser nanofabrication at resolutions as high as tens of nanometers.[1-7] The high resolutions achieved by TPP result from a combined contribution of the use of high numerical aperture (NA) objectives and existence of a threshold intensity in initiating the photopolymerization process. The choice of high NA objectives of short working distances limits the heights of end products of TPP, which has recently been circumvented by replacing the conventional focusing scheme with a simultaneous spatiotemporal focusing (SSTF) scheme.[8,9] The general concept of SSTF was proposed by Zhu et al for demonstrating wide-field 3D bioimaging based on two-photon fluorescence excitation.[10] However, for femtosecond laser 3D direct writing, the optical layout of SSTF must be changed for creating a tight spherical focal spot rather than a thin focal plane as shown by He et al.[11] The working mechanism of generation of a tight SSTF spot is briefly given as follow. In the SSTF, the femtosecond pulse first passes through a pair of gratings, thus the different frequency components in the pulse can be spatially separated from each other before they incident upon the back aperture of the focal lens. This reduces the local bandwidth of the pulse and leads to a dramatic extension of the pulse width. Temporal focusing, or focusing in the time domain, occurs after the focal lens during propagation of the pulse toward the geometric focal point because all the frequency components tend to recombine at the focus. In this way, the initial transform-limited pulse transiently restores itself at the focus while being stretched either before or after the focus, enabling shortening of the focal depth in terms of the peak intensity.[12,13]

Here, we show that when combining the SSTF with TPP, the nonlinear threshold effect in TPP uniquely allows for producing multi-scale 3D structures with a variable fabrication resolution in an alignment-free manner. This is benefitted from the fact that the size of laser affected volume with the STTF spot linearly depends on the power of the laser beam, making it extremely easy to dynamically tune the fabrication resolution during the continuous writing process. This provides an efficient means for boosting the throughput in fabricating multi-scale 3D microstructures, as otherwise the entire structures must be fabricated at the highest resolution (i.e., corresponding to a low fabrication throughput) as required by the regions of finest features.

**2. Experiment Section**

Our experimental setup is schematically illustrated in **Figure 1**. The positively pre-chirped femtosecond laser pulses delivered by a commercial laser amplifier (Libra, Coherent, Inc.) were of a bandwidth of ~27 nm and a pulse duration of ~ 200 ps. The pulse energy was tuned by a half waveplate combined with a polarizer. After reducing the beam width by a telescope system, the spectral components of the femtosecond laser pulses were first spatially separated by passing through a single-pass grating compressor which consists of two 1500 grooves/mm gratings blazed for the incident angle of 53°. The distance between the grating pair was adjusted to ~730 mm to compensate for the initial positive temporal dispersion of the beam, and the spatial chirp of the spatially separated frequency components was held constant at ~0.75 nm/mm after the grating pair. Then the spatially chirped pulses were focused into a SU8 resin by an objective lens (Leica 2 ×, NA =0.35). In combination with the threshold effect of the TPP, SSTF have the capability to generate the spherical voxel within the photosensitive resins.

**3. Dependence of fabrication resolution on the laser power**

To demonstrate the 3D isotropic fabrication resolution offered by the SSTF-TPP, we first fabricated two arrays of rods, one of which oriented along X direction while the other along Y direction, both embedded in a cube of SU-8 resin as shown in **Figure 2**a. The cube was intentionally fabricated to mechanically stabilize the logpile structures for the subsequent examination. The inset of Figure 2a shows the scanning electron microscopy (SEM) of the fabricated structures. All the rods were written at a fixed scan speed of 400 μm/s, whilst the laser power was varied to write the rods of different cross sectional sizes in the two arrays. It should be noted that the effective TPP printing speed depends on various parameters, such as the repetition rate of the femtosecond laser, the laser power and the resolution chosen for the fabrication. In our work, the major limit of the scanning speed was the low repetition rate of our femtosecond laser which is only 1 kHz. With this low repetition rate, the scan speed was kept below 400 μm/s to ensure sufficient overlap between consecutive laser shots. However, because of the enlarged focal spot with SSTF, the writing speed was 40 times higher than that in previous work which also employed a 1 kHz femtosecond laser.[14]

Figure 2b-2g shows the optical micrographs of cross sections of the rods fabricated at different femtosecond laser powers. The laser powers below were measured before the grating pair. Considering a total loss of 35% due to the two reflections at the gratings, the laser powers would drop to 65% of the measured powers at the back aperture of the objective lens. In Figure 2b, the diameters of the rods written with a laser power of 1.5 mW were measured to be ~9.4 μm for both X and Y scan directions, providing a direct evidence on the isotropic fabrication resolution. Then, we gradually raised the laser power to 2 mW, 3 mW and 4 mW in writing each pair of the rods oriented perpendicular to each other, i.e., one in X direction and the other in Y direction. The cross sectional micrographs of the rod pairs fabricated at the increasing laser power are respectively presented in Figure 2c to 2e. For each pair of the rods,

a nearly circular cross sectional shape has been achieved, resulting in various 3D isotropic resolutions of ~13.5 μm, ~20.5 μm and ~27 μm. When the laser power further increased to 5 mW and 6 mW, the cross sections of the fabricated rods slightly deviated from the perfect circular shape, i.e., the rods exhibit diamond-shaped cross sections as shown in Figure 2f and 2g. At the laser powers of 5 mW and 6 mW, the respective lateral sizes of the rods were measured to be 31.6 μm and 37.7 μm, while the longitudinal sizes were measured to be slightly larger, i.e., 34.5 μm and 44 μm, respectively.

To quantitatively determine the dependence of the fabrication resolution on the power of femtosecond laser, the lateral and longitudinal cross sectional sizes of the rods are plotted as functions of laser power in **Figure 2h**. We observe that in the range of 1.5 mW and 6 mW, the lateral and longitudinal resolutions are well balanced, and the resolutions in both lateral and longitudinal directions show nearly linear dependence on the power of femtosecond laser. The mechanism behind the power dependence of the fabrication resolution is the threshold effect in the interaction of femtosecond laser pulses with transparent materials.[5]

**4. Application of the SSTF-TPP for 3D laser printing**

Freestanding 3D structures are ideal model structures to examine the alignment-free tunability in the fabrication resolution provided by the SSTF-TPP in a straightforward way. Therefore, we fabricated a series of coils in SU-8 resin at different laser powers, as shown in **Figure 3**. In the writing process as illustrated in Figure 3a, we scanned the SSTF focal spot along a circular trajectory in XY plane while the motion stage was translated along Z direction. Figure 3b to 3d show the optical micrographs of the coils fabricated at various laser power of 3 mW, 4 mW, and 5 mW, respectively. The coils all have a same diameter of 200 μm in XY plane, but they are of different pitches in Z direction. The pitch of the coils in Figure 3b and 3C is 250 µm, while the pitch of the coil in Figure 3d is 150 µm. The coils show smooth surfaces and circular cross sections by examining at different angles of view. This is consistent with the results in Figure 2 and provides another unambiguous evidence on the isotropic 3D fabrication resolution achieved with SSTF-TTP.

Furthermore, we fabricated a multi-scale 3D structures in a single continuous writing process. **Figure 4**a shows the design of the structure, which is composed of a boat of a height of 8 mm, an elephant of a height of 3 mm standing on the deck of the boat, and a dog of a height of 1 mm sitting on top of the cabin. Because of the different feature sizes of the boat, the elephant, and the dog, these parts can be fabricated at different fabrication resolutions. In the experiment, we chose the contour scan method to fabricate the 3D structure for its relatively high fabrication efficiency. The boat was sliced into 200 layers with a layer height of 40 µm and fabricated at the laser power of 6 mW. Meanwhile, the elephant was sliced into 167 layers with a layer height of 18 µm and fabricated at the laser power of 3 mW, and the dog was sliced into 125 layers with a layer height of 8 µm and fabricated at the laser power of 1 mW. The parameters chosen above give rise to the fabrication resolutions of 40 µm, 18 µm, and 8 µm in writing of the boat, the elephant, and the dog, respectively. Figure 4b, 4c, and 4d present the digital-camera-captured images of a top view of the fabricated structure, one side view with the face of the elephant, and the opposite side view with the back of the elephant, respectively.

The quality of the fabricated sculptures of the dog and elephant was further examined with a scanning electron microscope (SEM), as shown in **Figure 5**. Figure 5a presents the overview of both the dog and elephant, showing their different feature sizes. Figure 5b shows that the details in the fabricated elephant such as the ear, the tusk, the trunk, the tail, and even the backbone can all be clearly seen. Likewise, Figure 5c shows the details in the fabricated dog, which are of a high quality as well thanks to the 8 μm resolution chosen for writing it.

## 5. Conclusion

To conclude, we demonstrate fabrication of multi-scale 3D structures using the SSTF-based TPP. We show that the fabrication resolution, which is isotropic in all the three dimensions in space, can be dynamically tuned only by varying the laser power. This is important for real-world applications in which it is preferable to avoid the cumbersome and time-consuming processes of precise realignment of the optics for resetting the fabrication resolution.

In the current experiment which employed a Ti:Sapphire laser of a low repetition rate of 1 kHz, it took around 10 hrs to complete the whole fabrication process for the structure shown in Fig. 4(a). However, the fabrication duration can be significantly shortened by replacing the 1 kHz laser source with high repetition rate, high average power laser systems. Moreover, the demonstrated tuning range of the fabrication resolution achievable by merely using the threshold effect is between sub-10 μm and ~40 μm. Meanwhile, the tuning range can be easily expanded to reach either higher or lower resolutions by modifying the focusing paramters, i.e., synergetically adjusting the width of the incident femtosecond laser beam and the distance between the two gratings, or replacing the focal lenses of different NAs. The wide tuning range of fabrication resolution renders the SSTF-TPP attractive for multi-scale 3D printing with unprecedented flexibility. Thus, the sophisticated capability of balancing the fabrication throughput and the fabrication resolution in the SSTF-TPP laser printing will benefit a number of applications such as bioengineering, microelectromechanics (MEMS), photonics, microfluidics, and manufacturing, to name a few.

**Acknowledgements**

This work is supported by the National Basic Research Program of China (Grant 2014CB921303), National Natural Science Foundation of China (Grants 11734009, 11674340, 61327902, 61590934, 61675220, 61505231), Strategic Priority Research Program of the Chinese Academy of Sciences (Grant XDB16030300), Key Research Program of Frontier Sciences, Chinese Academy of Sciences (Grant QYZDJ-SSW-SLH010), Project of Shanghai Committee of Science and Technology (Grant 17JC1400400) and Shanghai Rising-Star Program (Grant 17QA1404600).

Received: ((will be filled in by the editorial staff))
Revised: ((will be filled in by the editorial staff))
Published online: ((will be filled in by the editorial staff))

**Keywords**: (femtosecond laser two photon polymerization, multiscale 3D printing, simultaneous spatiotemporal focusing)

**Figure 1**. Schematic of the experimental setup. HWP, half waveplate; PBS, polarizing beam-splitter; L1, L2, telescope system; M, reflective mirrors; G1, G2, gratings; DM, dichroic mirror. Inset: The evolution of the pulse duration when the light propagated in focal volume.

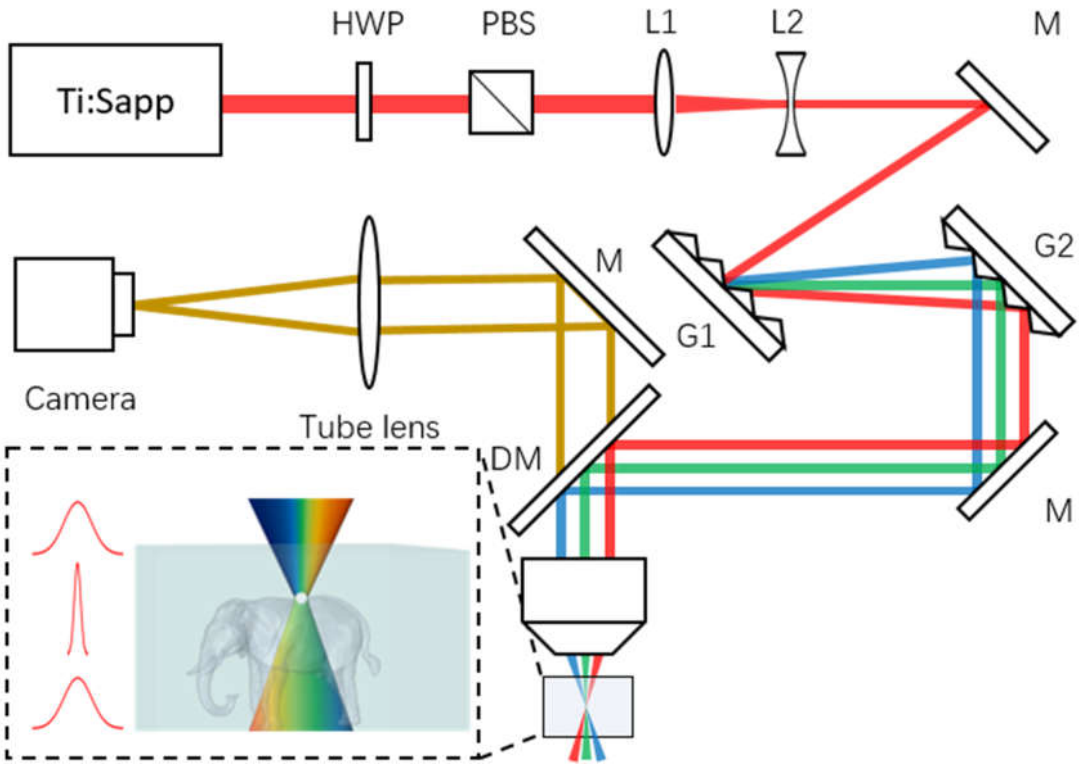

**Figure 2.** Control of 3D isotropic resolution. (a) Illustration of the designed model structure for measuring the fabrication resolutions at different femtosecond laser powers. Inset: SEM images of the fabricated rods with laser power of 3 mW (upper) and 5 mW (lower). (b-g): the end-sectional-view optical micrographs in XZ (left column) and YZ (right column) with the laser power of 1.5 mW, 2 mW, 3 mW, 4 mW, 5 mW, and 6 mW, respectively. (h): the lateral and longitudinal diameters of the rods plotted as functions of laser power.

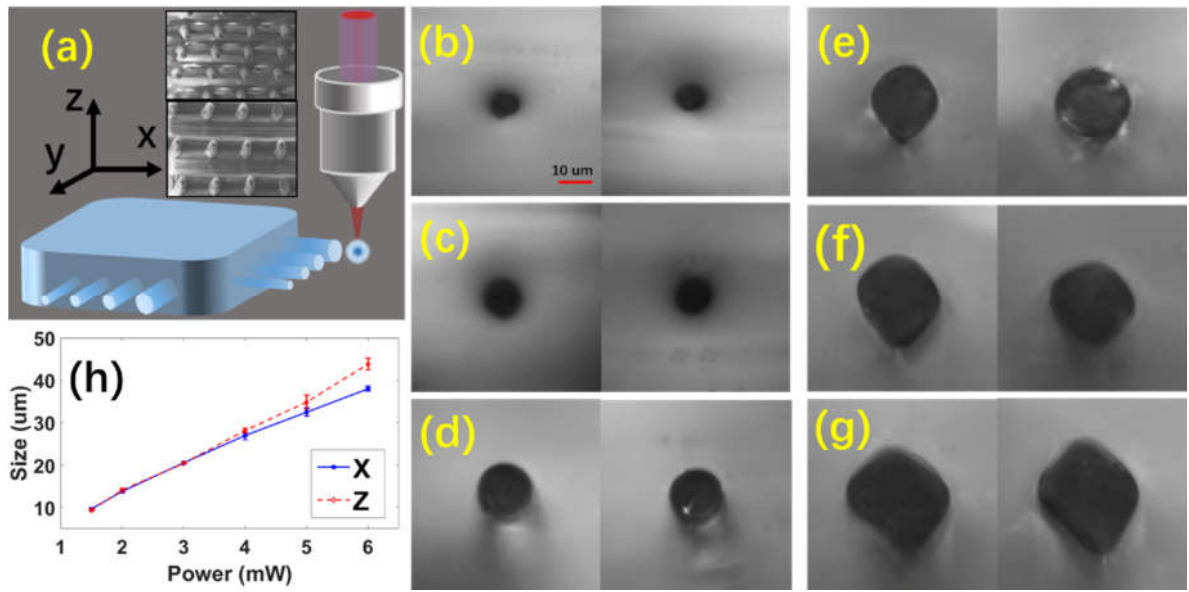

**Figure 3.** (a) Illustration of the coils fabrication, and the optical micrographs of the fabricated coils produced at different laser powers of (b) 3 mW, (c) 4 mW, and (d) 5 mW. Inset: close-up views of the cross sections at the end of the coils.

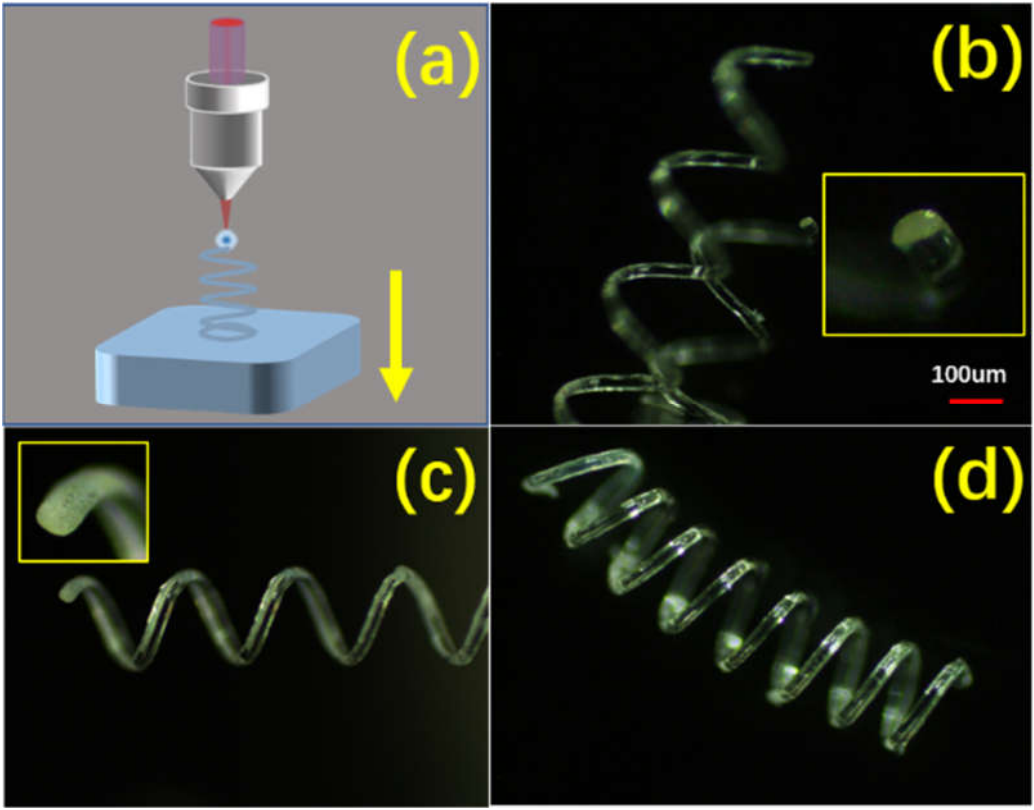

**Figure 4.** (a) Illustration of a model 3D structure, and fabricated structure shown by (b) the top view, (c) the side view with the face of the elephant, and (d) the side view with the back of the elephant.

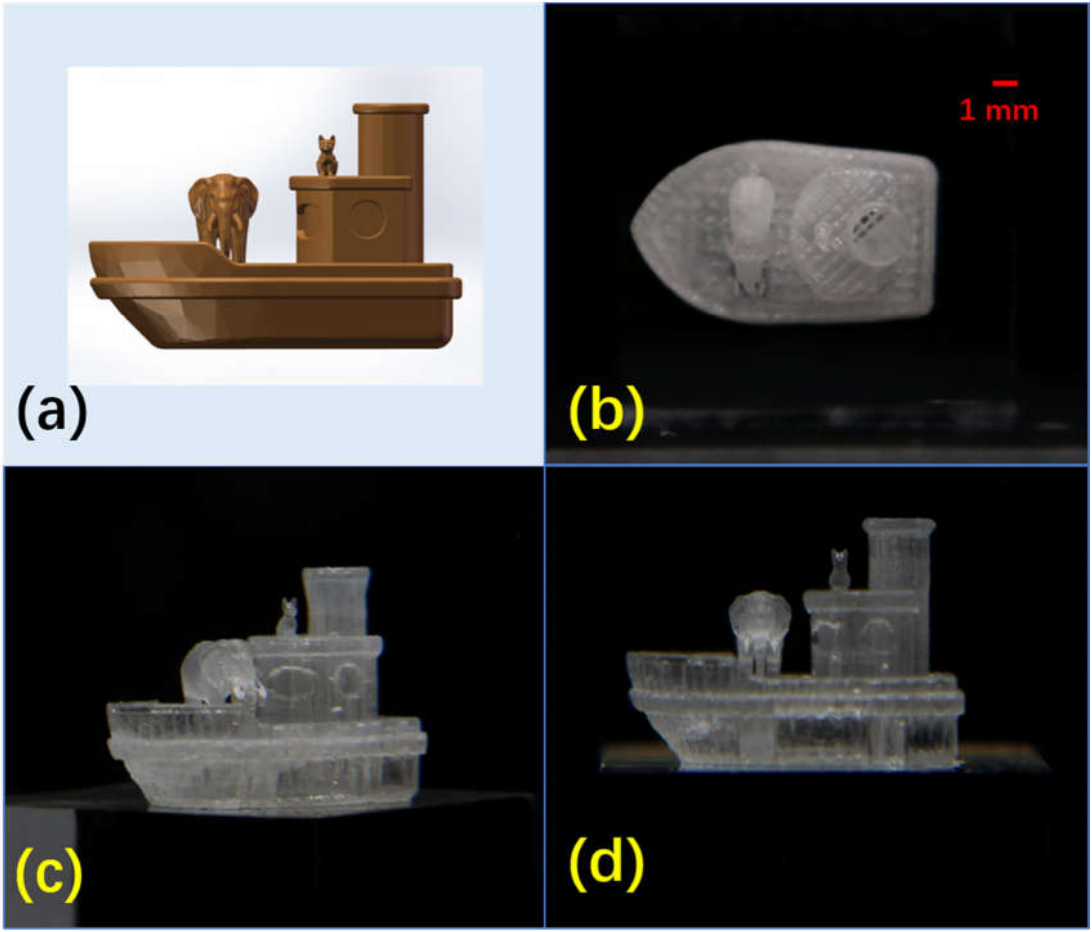

**Figure 5.** SEM images of the animals on the boat. (a) The overview of both the dog and the elephant. (b) The top view of the elephant. (c) The side view of the dog.

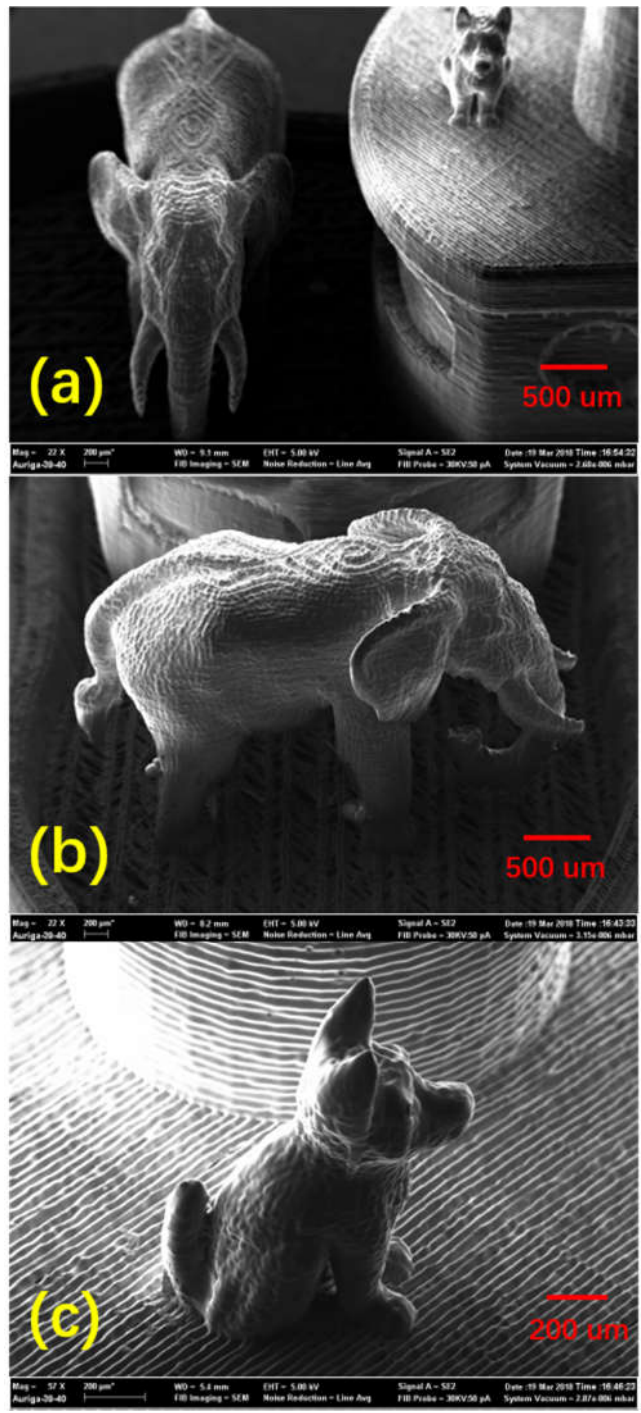